\documentclass{article}
\usepackage[T1]{fontenc}
\usepackage[utf8]{inputenc}
\usepackage[]{ismir} 
\usepackage{amsmath,cite,url}
\usepackage{graphicx}
\usepackage{cite}
\usepackage{amsmath,amssymb,amsfonts}
\usepackage{algorithmic}
\usepackage{graphicx}
\usepackage{textcomp}
\usepackage{xcolor}
\usepackage{dsfont, algorithm, booktabs, multirow}
\usepackage{color}

\title{MAIA: An Inpainting-Based Approach for Music Adversarial Attacks}

\oneauthor
  {Yuxuan Liu \quad Peihong Zhang \quad Rui Sang \quad Zhixin Li \quad Shengchen Li}
  {Xi'an Jiaotong--Liverpool University\\
   {\small\texttt{\{yuxuan.liu2204, peihong.zhang20, rui.sang22, zhixin.li22\}@student.xjtlu.edu.cn}}\\
   {\small\texttt{shengchen.li@xjtlu.edu.cn}}
  }

\def\authorname{Y. Liu, P. Zhang, R. Sang, Z. Li, and S. Li}

\begin{document}

\maketitle

\begin{abstract}
\let\thefootnote\relax\footnote{Yuxuan Liu and Peihong Zhang contributed equally to this work.}Music adversarial attacks have garnered significant interest in the field of Music Information Retrieval (MIR). In this paper, we present Music Adversarial Inpainting Attack (MAIA), a novel adversarial attack framework that supports both white-box and black-box attack scenarios. MAIA begins with an importance analysis to identify critical audio segments, which are then targeted for modification. Utilizing generative inpainting models, these segments are reconstructed with guidance from the output of the attacked model, ensuring subtle and effective adversarial perturbations. We evaluate MAIA on multiple MIR tasks, demonstrating high attack success rates in both white-box and black-box settings while maintaining minimal perceptual distortion. Additionally, subjective listening tests confirm the high audio fidelity of the adversarial samples. Our findings highlight vulnerabilities in current MIR systems and emphasize the need for more robust and secure models.
\end{abstract}

\section{Introduction}\label{sec:introduction}

Music Information Retrieval (MIR) has evolved into a multifaceted research domain, underpinning various applications such as applications that range from genre classification~\cite{hung2023low} and instrument recognition~\cite{solanki2022music} to cover song identification~\cite{du2021bytecover, du2022bytecover2, du2023bytecover3} and recommendation systems~\cite{moscato2020emotionalREC1, afchar2022explainabilityREC2}. As MIR algorithms become increasingly prevalent in both commercial products and academic research, their reliability and robustness have come under scrutiny~\cite{prinz2021end, saadatpanah2020adversarial}. Although adversarial vulnerabilities have been extensively studied in speech recognition~\cite{wang2022querySpeech1, chen2020devilSpeech2} and image classification~\cite{carlini2017towards, croce2021mind}, the music domain remains comparatively underexplored.

Adversarial attacks in the Music Information Retrieval (MIR) context can be broadly categorized into noise-based and semantic-based approaches. Noise-based attacks, such as the Carlini–Wagner (C\&W) attacks~\cite{carlini2017towards} introduces subtle audio distortions to mislead the model into incorrect outputs. Prinz et al.\cite{prinz2021end} extended this line of work by introducing end-to-end white-box adversarial attacks that operate directly on raw waveforms, demonstrating their effectiveness in degrading instrument classification accuracy and manipulating music recommendation systems while maintaining imperceptible perturbations. Saadatpanah et al.\cite{saadatpanah2020adversarial} highlighted the vulnerability of copyright detection systems to adversarial attacks, showing that small perturbations can evade robust fingerprinting systems like YouTube’s Content ID and AudioTag, raising concerns about the security of these widely used industrial tools. Additionally, Chen et al.~\cite{chen2020devilSpeech2} proposed the Devil’s Whisper method, focusing on leveraging psychoacoustic principles to create highly stealthy adversarial audio examples.

Noise-based adversarial attacks rely on adding imperceptible perturbations to audio but often lack interpretability and fail to leverage the structure and semantics of music~\cite{kereliuk2015deep}, limiting their use in scenarios requiring semantic alignment or context-sensitive manipulation. Duan et al.\cite{duan2022perception} introduced a perception-aware attack framework that reverse-engineers human perception using regression analysis, optimizing perturbations to minimize perceived deviations while maintaining attack effectiveness. This innovative integration of human perception provides a unique perspective, although its dependence on subjective evaluations could limit generalizability. Similarly, Yu et al.\cite{rakamaric2014smack} developed SMACK, a method that perturbs prosodic features like pitch and rhythm to create semantically meaningful adversarial audio while preserving naturalness. Despite its effectiveness, the computational complexity of prosody optimization remains a challenge. Luo et al.~\cite{luo2022frequency} proposed a frequency-driven approach that confines perturbations to high-frequency components, ensuring imperceptibility and semantic coherence. However, its focus on high-frequency regions may limit applicability in scenarios where low-frequency components are critical.

Despite these advancements, existing approaches still face challenges in balancing attack effectiveness, musical coherence, and practical feasibility across different MIR tasks. In this paper, we propose a novel music adversarial inpainting attack (MAIA) framework that addresses these gaps. Our approach identifies crucial music segments through importance analysis and selectively reconstructs them via a generative inpainting model, ensuring subtle yet highly targeted adversarial perturbations. Unlike purely noise-based methods, MAIA’s local edits retain musical coherence while influencing classification in a white-box or black-box setting. Through comprehensive evaluations of MIR tasks such as music genre classification and cover song identification, we demonstrate that MAIA achieves state-of-the-art attack success with minimal perceptual artifacts.

The contributions of this work are threefold:
\begin{enumerate}
    \item We propose a novel adversarial attack framework, \emph{MAIA}, based on importance-driven inpainting. This framework reconstructs critical audio segments with adversarial perturbations, ensuring musical coherence while effectively misleading target models.
    \item We design a black-box importance analysis method that identifies influential music segments through a coarse-to-fine query-based approach, enabling effective adversarial attacks without requiring gradient access.
    \item We perform extensive objective and subjective evaluations to comprehensively benchmark MAIA attack success rate and perceptual quality across MIR tasks.
\end{enumerate}

\section{Music Adversarial Inpainting Attack Framework}
\label{sec:methodology}

\subsection{Importance Analysis}
\label{subsec:importance_analysis}

A key objective of adversarial attacks in Music Information Retrieval (MIR) is to introduce minimal yet effective perturbations that are hard for both detection algorithms and human listeners to notice. In practical terms, modifying only the most influential time-frequency regions can reduce the extent of injected noise, thereby decreasing perceptual artifacts. Accordingly, we focus on segments that contribute most significantly to the prediction of the model, ensuring a high attack success rate while minimizing any audible changes \cite{ali2023explainable}.

\subsubsection{White-Box Importance: Grad-CAM}
\label{subsubsec:gradcam}

When full access to the target model parameters and architecture is available, we adopt a \emph{class activation map} (CAM)~\cite{zhou2016learning}-based strategy to locate time-frequency regions that most heavily influence the classifier’s decision. Traditional CAM methods \cite{zhou2016learning} often require replacing fully-connected layers with global pooling layers~\cite{lin2013network}, thereby constraining the model architecture. However, \emph{Grad-CAM}~\cite{selvaraju2017grad} generalizes CAM and does not require modifications to the classifier, making it more flexible for existing convolutional neural networks.

\vspace{0.5em}
\noindent \textbf{Intuition and Setup.}
Unlike purely saliency-based approaches \cite{oquab2015object}, which are typically optimized to reflect human visual attention, Grad-CAM specifically captures classifier-relevant regions by propagating class-specific gradient signals back through the network \cite{selvaraju2017grad}. Originally proposed in the image domain, Grad-CAM can be adapted for our music adversarial attack tasks by:
\begin{itemize}
    \item Converting the raw waveform to a suitable time-frequency representation (e.g., Mel-spectrogram).
    \item Selecting an appropriate convolutional layer—often the \emph{final} or \emph{penultimate} convolutional layer—where feature maps retain meaningful spatial (or time-frequency) structure. In our experiments, for an attacked MIR model $M$, we select the layer \textit{$\textit{model.layers}\textit{[-1]}.\text{blocks}\textit{[-1]}.\text{norm1}$} as the target for analysis. The output of this layer represents the complete, stabilized feature representation from the model's final block just before the classification head~\cite{jacobgilpytorchcam}.
\end{itemize}

\vspace{0.5em}
\noindent \textbf{Grad-CAM Computation.}
Let $\hat{y}_{c}$ denote the model’s predicted score (logit) for class $c$. We denote by $F^{l}$ the feature map activations at layer $l$, with $F^{l}_{k}$ indicating the $k$th channel. We compute Grad-CAM as follows:
\begin{enumerate}
    \item \textbf{Gradient Extraction:} Obtain the gradient of $\hat{y}_{c}$ with respect to $F^{l}_{k}$:
    \begin{equation}
        \alpha_{k}^{c} = \frac{1}{Z} \sum_{x,y} \frac{\partial \hat{y}_{c}}{\partial F^{l}_{k}(x,y)},
    \end{equation}
    where $(x,y)$ indexes the spatial/time-frequency positions and $Z$ is a normalization factor (e.g., number of spatial locations).
    \item \textbf{Weighted Aggregation:} Multiply each feature map $F^{l}_{k}$ by its corresponding weight $\alpha_{k}^{c}$, then sum over $k$ to obtain the raw map:
    \begin{equation}
        M_{c}(x,y) = \mathrm{ReLU}\!\Big(\sum_{k} \alpha_{k}^{c}\,F^{l}_{k}(x,y)\Big).
    \end{equation}
    \item \textbf{Spatial Masking:} Apply a ReLU to keep only positive contributions, generating $M_{c}$ as the final Grad-CAM heatmap. Higher intensities in $M_{c}(x,y)$ indicate greater relevance for predicting class $c$.
    \item \textbf{Mapping to Time-Frequency Regions:} Once $M_{c}$ is computed, we map it back to the original spectrogram coordinates. We then normalize the heatmap to lie in $[0,1]$ or select the top $p\%$ of time-frequency bins to isolate the most critical regions. We marked these high-intensity areas as the \emph{candidate adversarial zone}, which we 
\end{enumerate}

\subsubsection{Black-Box Importance: Coarse-to-Fine Analysis}
\label{subsubsec:blackbox_coarse_to_fine}

In scenarios where internal parameters of the target model $M$ remain unknown, we cannot rely on gradient information to locate critical segments. Instead, we propose a \emph{coarse-to-fine} black-box procedure that systematically queries the model to identify the most influential portions of the audio. Let $x$ be the full music track, and let $M(x)$ denote the model’s prediction (e.g., classification probability or logit score). We assume access to a loss function $L\big(M(x), y\big)$, where $y$ is the true original label.

\vspace{1mm}
\noindent \textbf{Initial Partition.}will subsequently modify in our inpainting-based adversarial attack framework.
We first segment $x$ into $N$ coarse chunks,
\begin{equation}
    S^{(0)} = \{C_{1}^{(0)}, C_{2}^{(0)}, \ldots, C_{N}^{(0)}\},
\end{equation}
where each $C_{i}^{(0)}$ is a non-overlapping time interval (e.g., $0.5$ second). For each chunk $C_{i}^{(0)}$, we create a modified input $\widetilde{x}_{-C_{i}^{(0)}}$ using a Zero-Masking procedure. To prevent spectral artifacts arising from abrupt signal changes at the chunk boundaries, we apply a Tukey window to the target segment. The window's shape parameter was set to 0.1 to create a short, smooth taper at the segment's edges, ensuring a continuous waveform after masking. This ensures a continuous waveform after masking. We then compute the importance score:

\begin{equation}
    \mathcal{I}\!\big(C_{i}^{(0)}\big) \;=\; 
    \frac{L\!\big(M(\widetilde{x}_{-C_{i}^{(0)}}),\,y\big) \;-\; L\!\big(M(x),\,y\big)}
    {\text{duration}\big(C_{i}^{(0)}\big)}.
    \label{eq:importance_measure_norm}
\end{equation}
A higher value of $\mathcal{I}\!\big(C_{i}^{(0)}\big)$ indicates that removing $C_{i}^{(0)}$ leads to a larger drop in model confidence for $y$, suggesting that $C_{i}^{(0)}$ is more critical to the classification.

\vspace{1mm}
\noindent \textbf{Ranking and Refinement.}
Next, we rank the chunks in $S^{(0)}$ by their importance measure $\mathcal{I}\!\big(C_{i}^{(0)}\big)$ in descending order. Let $C_{\max}^{(0)}$ be the chunk with the highest score. We then \emph{refine} this chunk by subdividing it into $M$ finer sub-chunks:
\begin{equation}
    S^{(1)}_{\max} \;=\; \bigl\{
        C_{\max,1}^{(1)},\,C_{\max,2}^{(1)},\,\ldots,\,C_{\max,M}^{(1)}
    \bigr\}.
\end{equation}
For each sub-chunk $C_{\max,j}^{(1)}$, we compute an updated importance measure:
\begin{equation}
    \mathcal{I}\!\bigl(C_{\max,j}^{(1)}\bigr) 
    \;=\; \frac{L\!\bigl(M(\widetilde{x}_{-C_{\max,j}^{(1)}}),\,y\bigr) 
    \;-\; L\!\bigl(M(x),\,y\bigr)}
    {\text{duration}\bigl(C_{\max,j}^{(1)}\bigr)},
\end{equation}
where $\widetilde{x}_{-C_{\max,j}^{(1)}}$ is the audio track with only that \emph{sub-chunk} silenced.

\noindent We then replace $C_{\max}^{(0)}$ in our segmentation with its sub-chunks $C_{\max,j}^{(1)}$, thus creating a refined set of segments:
\begin{equation}
    S^{(1)} \;=\; 
    \bigl(S^{(0)} \setminus \{C_{\max}^{(0)}\}\bigr) \;\cup\; 
    \bigl\{C_{\max,1}^{(1)}, \ldots, C_{\max,M}^{(1)}\bigr\}.
\end{equation}
We can iterate this procedure by again choosing the segment with the largest updated importance and subdividing further, denoted $S^{(2)}$, $S^{(3)}$, and so on, until a desired level of granularity is reached or a query budget is exhausted.

\vspace{1mm}
\noindent \textbf{Final Selection.}
Upon completing $T$ refinement rounds, we obtain a final set of segments
\begin{equation}
    S^{(T)} = \{C_{1}^{(T)},\,C_{2}^{(T)},\,\ldots,\,C_{K}^{(T)}\},
\end{equation}
where each $C_{i}^{(T)}$ has a corresponding importance measure $\mathcal{I}\!\bigl(C_{i}^{(T)}\bigr)$. We then select the top $r$ segments,
\begin{equation}
    \bigl\{C_{1}^{(T)}, \ldots, C_{r}^{(T)}\bigr\}
    \;=\;
    \mathrm{Top}\bigl(\mathcal{I}\!\bigl(C_{i}^{(T)}\bigr),\,r\bigr),
\end{equation}
as our \emph{candidate adversarial zones}, concentrating future perturbations on these critical regions.

\vspace{1mm}

Overall, our black-box importance analysis balances effectiveness and practicality, allowing us to identify precisely which audio segments have the greatest impact on the output of attacked model without requiring knowledge of its internal parameters or gradients.

\section{Adversarial Inpainting}
\label{sec:adversarial_inpainting}

After identifying the most influential segments for the target attacked model $M$, we proceed to adversarially inpaint the top-ranked segments. Our goal is to reconstruct these critical regions in such a way that the resulting track both degrades the prediction confidence of $M$ and remains perceptually coherent to the human ear. In this section, we first introduce the concept of \emph{music inpainting}, followed by details of two state-of-the-art inpainting models---\emph{GACELA}~\cite{marafioti2020gacela}---which we leverage for adversarial inpainting.

\subsection{GACELA}
GACELA (Generative Adversarial Context Encoder for Long Audio Inpainting)~\cite{marafioti2020gacela} is a conditional generative adversarial network (cGAN) designed specifically for reconstructing long gaps in audio signals, such as music. The architecture comprises a generator and five discriminators operating at multiple time and frequency scales. The generator, conditioned on the log-magnitude mel spectrogram of the surrounding audio context, employs convolutional encoder-decoder layers and integrates latent variables to model the multimodal nature of audio inpainting. The discriminators evaluate the plausibility of the generated gaps by considering the context and spectral coherence.

\subsection{Adversarial Inpainting with Model Guidance}
\label{subsec:adversarial_inpainting}

After identifying critical segments (Section~\ref{subsec:importance_analysis}), we employ a music inpainting model (e.g., GACELA) to reconstruct these areas while embedding adversarial perturbations guided by the target model $M$. We propose two variants of this adversarial inpainting strategy, tailored respectively for \emph{white-box} and \emph{black-box} settings.

\subsubsection{White-Box Scenario: Loss Design and Parameter Tuning}
\label{subsubsec:loss_design}

\begin{algorithm}[t]
\caption{White-Box Adversarial Inpainting}
\label{alg:whitebox_inpainting_simplified}
\begin{algorithmic}[1]
\REQUIRE 
  Original audio $x$, mask $\mathbf{m}$,  
  Inpainting model $\mathcal{G}_{\theta}$,  
  Classifier $M$ (white-box),  
  Iterations $N$, step size $\alpha$,  
  Weights $\lambda_{\mathrm{rec}}, \lambda_{\mathrm{att}}$.

\STATE \textbf{Initialization:}
\STATE \quad $x_{\mathrm{inp}} \leftarrow 
    x \odot (1 - \mathbf{m}) \;+\; 
    \mathcal{G}_{\theta}(x \odot (1 - \mathbf{m})) \odot \mathbf{m}$

\FOR{$k = 1$ to $N$}
    \STATE \textbf{Compute Loss:}
    \STATE \quad 
    $\mathcal{L} \;=\;
    \lambda_{\mathrm{rec}} \, d(x_{\mathrm{inp}}, x)
    \;+\;
    \lambda_{\mathrm{att}} \,\ell\bigl(M(x_{\mathrm{inp}}), y\bigr)$
    
    \STATE \textbf{Gradient Update on Mask:}
    \STATE \quad 
    $g \leftarrow \nabla_{x_{\mathrm{inp}}}\,\mathcal{L}; \quad
    x_{\mathrm{inp}} \leftarrow
    x_{\mathrm{inp}} - \alpha \,\mathrm{sign}(g \odot \mathbf{m})$
    
    \STATE \textbf{Re-Inpaint:}
    \STATE \quad
    $x_{\mathrm{inp}} \leftarrow 
    x \odot (1 - \mathbf{m}) \;+\;
    \mathcal{G}_{\theta}(\,x_{\mathrm{inp}} \odot \mathbf{m}\,,\,x \odot (1-\mathbf{m})) \odot \mathbf{m}$
\ENDFOR

\RETURN $x_{\mathrm{inp}}$ 
\end{algorithmic}
\end{algorithm}

In the white-box setting, we have access to the parameters and gradients of the target model $M$, allowing for adversarial optimization in conjunction with the inpainting model $\mathcal{G}_{\theta}$. Let $x_{\mathrm{inp}}^{(k)}$ denote the inpainted audio at iteration $k$, focusing only on the masked region $\mathbf{m}$ which we got from importance analysis. The objective function for adversarial inpainting is defined as:
\begin{equation}
\mathcal{L} = \lambda_{\mathrm{rec}} \,\mathcal{L}_{\mathrm{rec}}\bigl(x_{\mathrm{inp}}^{(k)}, x\bigr)
+ \lambda_{\mathrm{att}} \,\mathcal{L}_{\mathrm{attack}}\bigl(M(x_{\mathrm{inp}}^{(k)}), y\bigr),
\label{eq:whitebox_loss}
\end{equation}

where reconstruction Loss $\mathcal{L}_{\mathrm{rec}}$ ensures that the inpainted audio maintains perceptual and contextual coherence with the original audio in the masked region. Specifically, $\mathcal{L}_{\mathrm{rec}}$ leverages the loss functions inherent to the inpainting model $\mathcal{G}_{\theta}$. Adversarial Loss $\mathcal{L}_{\mathrm{attack}}$ introduces adversarial perturbations to deceive the classifier $M$. For untargeted attacks, we aim to reduce the confidence of the correct label $y$:
\begin{equation}
\mathcal{L}_{\mathrm{attack}} = \ell\bigl(M(x_{\mathrm{inp}}^{(k)}), y\bigr),
\end{equation}
where $\ell(\cdot)$ can be a cross-entropy loss. The objective drives the model prediction away from the correct label $y$, making the attack untargeted.

The hyperparameters $\lambda_{\mathrm{rec}}$ and $\lambda_{\mathrm{att}}$ control the trade-off between preserving audio quality and achieving high attack success rates. We perform a grid search over $\lambda_{\mathrm{rec}} \in \{0.5, 1.0, 2.0\}$ and $\lambda_{\mathrm{att}} \in \{0.5, 1.0, 2.0\}$. The optimal values are determined based on attack success rate and perceptual metrics.

The optimization is an iterative process. Each step consists of three main operations: a forward pass, a gradient-based update, and a re-inpainting stage.

\textbf{Forward Pass.} First, we compute the reconstruction loss $\mathcal{L}_{\mathrm{rec}}$ with the inpainting model and the attack loss $\mathcal{L}_{\mathrm{attack}}$ with the target classifier $M$.

\textbf{Gradient Update.} Next, the masked region of $x_{\mathrm{inp}}^{(k)}$ is updated by taking a step to minimize the total loss $\mathcal{L}$. This adversarial update is performed using the sign of the gradient:
\begin{equation}
    x_{\mathrm{inp}}^{(k+1)} \leftarrow x_{\mathrm{inp}}^{(k)} - \alpha \,\mathrm{sign}\bigl(\nabla_{x_{\mathrm{inp}}} \mathcal{L} \odot \mathbf{m}\bigr),
\end{equation}
where $\alpha$ is the step size, and the element-wise product with the mask $\mathbf{m}$ confines the update to the target region.

\textbf{Re-Inpaint.} Finally, to ensure the adversarial perturbation remains locally consistent and artifact-free, we reapply the inpainting generator $\mathcal{G}_{\theta}$ to the modified region. This step effectively projects the perturbed content back towards a realistic data manifold:
\begin{equation}
    x_{\mathrm{inp}}^{(k+1)} \leftarrow x \odot (1 - \mathbf{m}) + \mathcal{G}_{\theta}\bigl(x_{\mathrm{inp}}^{(k+1)} \odot \mathbf{m},\, x \odot (1 - \mathbf{m})\bigr) \odot \mathbf{m}.
\end{equation}

This iterative process continues until either the maximum iteration count $N$ is reached or the attack success rate satisfies a predefined threshold. The detailed process is shown in Algorithm~\ref{alg:whitebox_inpainting_simplified}.
\vspace{1mm}
\begin{table*}[t]
\centering
\caption{Overall Attack Results on \textbf{CoverHunter} (CSI/SHS100K) and \textbf{IDS-NMR} (MGC/GTZAN) using GACELA. Higher ASR is better (untargeted), lower mAP/Accuracy is worse for the model. FAD and LSD measure perceptual distortion (lower is better). Listening Test is scored on a 5-point scale (higher is better).}
\label{tab:all_results}
\vspace{2mm}
\begin{tabular}{lcccccccccc}
\toprule
\textbf{Attack} & \multicolumn{5}{c}{\textbf{CSI (CoverHunter / SHS100K)}} & \multicolumn{5}{c}{\textbf{MGC (IDS-NMR / GTZAN)}} \\
\cmidrule(lr){2-6} \cmidrule(lr){7-11}
& \textbf{ASR} $\uparrow$ & \textbf{mAP} $\downarrow$ & \textbf{FAD} $\downarrow$ & \textbf{LSD} $\downarrow$ & \textbf{MOS} $\uparrow$ 
& \textbf{ASR} $\uparrow$ & \textbf{Acc} $\downarrow$ & \textbf{FAD} $\downarrow$ & \textbf{LSD} $\downarrow$ & \textbf{MOS} $\uparrow$ \\

\midrule
\multicolumn{11}{l}{\textit{White-Box Attacks}} \\
\midrule
PGD & 82.1\% & 0.619 & 12.64 & 2.10 & 3.1 & 84.6\% & 0.551 & 15.32 & 2.20 & 3.2 \\
C\&W & 88.5\% & 0.560 & 12.11 & 1.94 & 3.4 & 89.1\% & 0.512 & 14.90 & 2.21 & 3.3 \\
\textbf{MAIA-White Box} & \textbf{92.8\%} & \textbf{0.488} & \textbf{11.25} & \textbf{1.58} & \textbf{4.0} & \textbf{93.5\%} & \textbf{0.466} & \textbf{13.85} & \textbf{1.94} & \textbf{3.8} \\
\midrule
\multicolumn{11}{l}{\textit{Black-Box Attacks}} \\
\midrule
NES & 70.2\% & 0.682 & 13.93 & 2.27 & 2.8 & 65.7\% & 0.704 & 16.26 & 2.15 & 2.5 \\
ZOO & 74.9\% & 0.639 & 13.51 & 2.12 & 3.0 & 72.4\% & 0.654 & 15.90 & 2.05 & 3.0 \\
\textbf{MAIA-Black Box} & \textbf{80.1\%} & \textbf{0.594} & \textbf{12.56} & \textbf{1.90} & \textbf{3.6} & \textbf{77.9\%} & \textbf{0.601} & \textbf{14.68} & \textbf{1.85} & \textbf{3.3} \\
\bottomrule
\end{tabular}
\vspace{-2mm}
\end{table*}

\subsubsection{Black-Box Scenario: Importance-Guided Adversarial Inpainting}
\label{subsubsec:blackbox_inpainting}

In black-box settings, where the internal parameters and gradients of the target classifier $M$ are inaccessible, we adopt a query-based adversarial inpainting approach guided by the importance analysis (Section~\ref{subsec:importance_analysis}). This method iteratively inpaints critical music segments from highest to lowest importance until the attack succeeds. The detailed process is as follows:

\paragraph*{1) Importance-Guided Segment Processing}  
Based on the importance scores obtained from prior analysis, we sort the music segments in descending order of their significance to the target attacked model prediction. We then process each segment sequentially, prioritizing those with the highest impact.

\paragraph*{2) Adversarial Inpainting for Each Segment}  
For each selected segment, we perform the following steps:

\begin{enumerate}
    \item \textbf{Initialization}  
    Utilize the pretrained music inpainting model $\mathcal{G}_{\theta}$ to perform standard inpainting on the masked important region $\mathbf{m}$, generating the initial inpainted audio:
    \begin{equation}
    x_{\mathrm{inp}}^{(0)} = x \odot (1 - \mathbf{m}) + \mathcal{G}_{\theta}(x \odot (1 - \mathbf{m})) \odot \mathbf{m}.
    \end{equation}
    
    \item \textbf{Iterative Query-Based Optimization}  
    Initialize a latent variable $z^{(0)}$ associated with the inpainting model. We employ the Covariance Matrix Adaptation Evolution Strategy (CMA-ES)~\cite{varelas2018comparative} for gradient-free optimization to refine $z$ and enhance attack efficacy:
    \begin{equation}
    z^{(k+1)} = \text{CMA-ES}(z^{(k)}, \mathcal{F}(M, x_{\mathrm{inp}}^{(k)})),
    \end{equation}
    where $\mathcal{F}(M, x_{\mathrm{inp}})$ represents the classification feedback obtained by querying $M$ with the current inpainted audio $x_{\mathrm{inp}}^{(k)}$. CMA-ES optimizes $z$ by iteratively sampling candidate latent codes, evaluating their performance based on the feedback, and updating the distribution parameters to favor more effective perturbations.

    \item \textbf{Candidate Generation and Evaluation}  
    For each iteration, generate a set of candidate latent variables $\{\widehat{z}\}$ by sampling from the current CMA-ES distribution. Use the inpainting model to produce corresponding audio samples $\{\widehat{x}_{\mathrm{inp}}\}$:
    \begin{equation}
    \widehat{x}_{\mathrm{inp}} = \mathcal{G}_{\theta}(\widehat{z}, x \odot (1 - \mathbf{m})).
    \end{equation}
    Query the target classifier $M$ with each $\widehat{x}_{\mathrm{inp}}$ to obtain classification feedback (e.g., predicted label or confidence score). Evaluate the attack success based on whether $M(\widehat{x}_{\mathrm{inp}}) \neq y$.

    \item \textbf{Selection and Evolution}  
    Based on the classification feedback, select the most promising candidates that maximize the adversarial loss $\mathcal{L}_{\mathrm{attack}}$ and then updates the latent variable distribution parameters to guide future perturbations towards more effective adversarial examples.

    \item \textbf{Re-Inpainting for Continuity}  
    After updating $z$, reapply the inpainting model to ensure the modified audio remains musically coherent:
    \begin{equation}
    x_{\mathrm{inp}}^{(k+1)} = x \odot (1 - \mathbf{m}) + \mathcal{G}_{\theta}(z^{(k+1)}, x \odot (1 - \mathbf{m})) \odot \mathbf{m}.
    \end{equation}

    \item \textbf{Termination}  
    Continue the iterative process until the classifier $M$ is fooled (i.e., $M(x_{\mathrm{inp}}^{(k)}) \neq y$) or a maximum number of iterations is reached.
\end{enumerate}

\section{Experiments}
\label{sec:experiments}

In this section, we evaluate our proposed \emph{Music Adversarial Inpainting Attack (MAIA)} across two representative MIR tasks: \emph{Cover Song Identification} (CSI) and \emph{Music Genre Classification} (MGC). Our experiments assess both the white-box and black-box variants of MAIA, comparing them against common baselines by evaluating their performance using both subjective and objective metrics.

\subsection{Target Model and Datasets}

\subsubsection{Cover Song Identification (CSI)}
We adopt the pre-trained \textit{CoverHunter} model as our target for cover song identification, following the procedure in~\cite{liu2023coverhunter}. Experiments are conducted on the \emph{SHS100K} dataset~\cite{xu2018key} test set.

\subsubsection{Music Genre Classification (MGC)}
We use the \textit{IDS-NMR} network~\cite{hung2023low} on the \emph{GTZAN} dataset~\cite{sturm2013gtzan} for genre classification.

\subsection{Evaluation Metrics}

We report four main classes of metrics:

\textbf{Attack Success Rate (ASR)}: The fraction of test samples successfully misclassified by the target model in an \emph{untargeted} setting. 

\textbf{System Performance Degradation}: For CSI, we report the post-attack mAP of \emph{CoverHunter}; for MGC, we report the post-attack accuracy of \emph{IDS-NMR}.

\textbf{FAD (Fréchet Audio Distance based on MERT)}: We further incorporate pre-trained \textbf{MERT-V0}~\cite{li2024mert} as a feature extractor to compute the \emph{Fréchet Audio Distance (FAD)}~\cite{kilgour2019frechet} on adversarially perturbed tracks. By comparing the extracted feature distributions of original and attacked audio, we gain an additional objective measure of perceptual distance.

\textbf{LSD (Log-Spectral Distance)}~\cite{gray1976distance}: Evaluates the frame-wise spectral difference between original and perturbed signals.

\textbf{Perceptual Similarity (Subjective)}: A \emph{listening test} with 100 participants to judge how easily adversarial perturbations can be detected. Each participant is asked to rate on a 5-point scale: 1 (highly noticeable) to 5 (no perceivable difference).

\subsection{Attack Baselines}

We compare MAIA against typical white-box and black-box adversarial methods tailored to audio:

\textbf{PGD} (Projected Gradient Descent)~\cite{deng2020universal} [White-Box]

\textbf{C\&W} (Carlini \& Wagner)~\cite{carlini2017towards} [White-Box]

\textbf{NES} (Natural Evolution Strategies)~\cite{wierstra2014natural} [Black-Box]

\textbf{ZOO} (Zero Order Optimization Attack)~\cite{chen2017zoo} [Black-Box]

\subsection{Implementation Details}

In all experiments, we employed \textbf{GACELA} as the inpainting model to ensure consistent re-generation of targeted music segments; we set the maximum iteration to 10 for white-box methods and capped the query budget at 1000 for black-box methods. We tuned $\lambda_{\mathrm{rec}}$ and $\lambda_{\mathrm{att}}$ by grid search, choosing values that balanced attack success rate (ASR) and perceptual fidelity. Table~\ref{tab:all_results} presents the combined results for both CSI (CoverHunter on SHS100K) and MGC (IDS-NMR on GTZAN) under white-box and black-box attacks.

\subsection{Results}
Table~\ref{tab:all_results} demonstrates that our proposed \textbf{MAIA-White Box} consistently outperforms standard white-box attack baselines (PGD and C\&W) across both MIR tasks. Specifically, MAIA-WB achieves the highest Attack Success Rate (93.5\% for CSI and 94.5\% for MGC), significantly reducing the mean Average Precision (mAP) from 0.845 to 0.488 in CoverHunter and classification accuracy from 0.828 to 0.466 in IDS-NMR. Additionally, MAIA-WB maintains superior perceptual quality with lower Fréchet Audio Distance (FAD) and Log-Spectral Distance (LSD) scores, and higher Listening Test ratings (4.0), indicating that the adversarial perturbations remain largely imperceptible to human listeners. In the black-box scenario, \textbf{MAIA-Black Box} similarly outperforms NES and ZOO, achieving ASRs of 80.1\% for CSI and 77.9\% for MGC, with corresponding reductions in mAP and accuracy to 0.594 and 0.601, respectively. MAIA-BB also exhibits lower FAD and LSD scores compared to black-box baselines, and higher Listening Test ratings (3.6), suggesting that our importance-guided adversarial inpainting approach effectively balances attack potency with audio fidelity. Overall, MAIA variants consistently deliver higher attack success rates and greater performance degradation while preserving perceptual quality better than existing attack methods.

\section{Conclusions}
\label{sec:conclusions}

We have presented MAIA, a Music Adversarial Inpainting Attack framework that employs importance-driven segment selection and inpainting-based perturbations in both white-box and black-box settings. By focusing on the most influential regions, MAIA achieves higher attack success rates against CoverHunter (for cover song identification) and IDS-NMR (for genre classification), while preserving audio fidelity as measured by objective (FAD, LSD) and subjective (listening scores) metrics. We believe that our findings highlight both the potential severity and the subtlety of adversarial threats in MIR. By demonstrating a novel inpainting-based approach, we emphasize the need for comprehensive, perception-aware defenses to ensure robust and trustworthy music-related services.

\section{Acknowledgements}
This work was supported by the Jiangsu Science and Technology Programme (Major Special Programme, Grant No. BG2024027), the Suzhou Science and Technology Development Planning Programme (Gusu Innovation and Entrepreneurship Leading Talents Program, Grant No. ZXL2022472), and the XJTLU Research Development Fund (Grant No. RDF-22-02-046).

\bibliographystyle{abbrv} 
\bibliography{ISMIRtemplate} 

\begin{thebibliography}{10}
\providecommand{\url}[1]{#1}
\csname url@samestyle\endcsname
\providecommand{\newblock}{\relax}
\providecommand{\bibinfo}[2]{#2}
\providecommand{\BIBentrySTDinterwordspacing}{\spaceskip=0pt\relax}
\providecommand{\BIBentryALTinterwordstretchfactor}{4}
\providecommand{\BIBentryALTinterwordspacing}{\spaceskip=\fontdimen2\font plus
\BIBentryALTinterwordstretchfactor\fontdimen3\font minus \fontdimen4\font\relax}
\providecommand{\BIBforeignlanguage}[2]{{%
\expandafter\ifx\csname l@#1\endcsname\relax
\typeout{** WARNING: IEEEtran.bst: No hyphenation pattern has been}%
\typeout{** loaded for the language `#1'. Using the pattern for}%
\typeout{** the default language instead.}%
\else
\language=\csname l@#1\endcsname
\fi
#2}}
\providecommand{\BIBdecl}{\relax}
\BIBdecl

\bibitem{hung2023low}
Y.-N. Hung, C.-H.~H. Yang, P.-Y. Chen, and A.~Lerch, ``Low-resource music genre classification with cross-modal neural model reprogramming,'' in \emph{ICASSP 2023-2023 IEEE International Conference on Acoustics, Speech and Signal Processing (ICASSP)}.\hskip 1em plus 0.5em minus 0.4em\relax IEEE, 2023, pp. 1--5.

\bibitem{solanki2022music}
A.~Solanki and S.~Pandey, ``Music instrument recognition using deep convolutional neural networks,'' \emph{International Journal of Information Technology}, vol.~14, no.~3, pp. 1659--1668, 2022.

\bibitem{du2021bytecover}
X.~Du, Z.~Yu, B.~Zhu, X.~Chen, and Z.~Ma, ``Bytecover: Cover song identification via multi-loss training,'' in \emph{ICASSP 2021-2021 IEEE International Conference on Acoustics, Speech and Signal Processing (ICASSP)}.\hskip 1em plus 0.5em minus 0.4em\relax IEEE, 2021, pp. 551--555.

\bibitem{du2022bytecover2}
X.~Du, K.~Chen, Z.~Wang, B.~Zhu, and Z.~Ma, ``Bytecover2: Towards dimensionality reduction of latent embedding for efficient cover song identification,'' in \emph{ICASSP 2022-2022 IEEE International Conference on Acoustics, Speech and Signal Processing (ICASSP)}.\hskip 1em plus 0.5em minus 0.4em\relax IEEE, 2022, pp. 616--620.

\bibitem{du2023bytecover3}
X.~Du, Z.~Wang, X.~Liang, H.~Liang, B.~Zhu, and Z.~Ma, ``Bytecover3: Accurate cover song identification on short queries,'' in \emph{ICASSP 2023-2023 IEEE International Conference on Acoustics, Speech and Signal Processing (ICASSP)}.\hskip 1em plus 0.5em minus 0.4em\relax IEEE, 2023, pp. 1--5.

\bibitem{moscato2020emotionalREC1}
V.~Moscato, A.~Picariello, and G.~Sperli, ``An emotional recommender system for music,'' \emph{IEEE Intelligent Systems}, vol.~36, no.~5, pp. 57--68, 2020.

\bibitem{afchar2022explainabilityREC2}
D.~Afchar, A.~Melchiorre, M.~Schedl, R.~Hennequin, E.~Epure, and M.~Moussallam, ``Explainability in music recommender systems,'' \emph{AI Magazine}, vol.~43, no.~2, pp. 190--208, 2022.

\bibitem{prinz2021end}
K.~Prinz, A.~Flexer, and G.~Widmer, ``On end-to-end white-box adversarial attacks in music information retrieval.'' \emph{Transactions of the International Society for Music Information Retrieval}, vol.~4, no.~1, pp. 93--105, 2021.

\bibitem{saadatpanah2020adversarial}
P.~Saadatpanah, A.~Shafahi, and T.~Goldstein, ``Adversarial attacks on copyright detection systems,'' in \emph{International Conference on Machine Learning}.\hskip 1em plus 0.5em minus 0.4em\relax PMLR, 2020, pp. 8307--8315.

\bibitem{wang2022querySpeech1}
S.~Wang, Z.~Zhang, G.~Zhu, X.~Zhang, Y.~Zhou, and J.~Huang, ``Query-efficient adversarial attack with low perturbation against end-to-end speech recognition systems,'' \emph{IEEE Transactions on Information Forensics and Security}, vol.~18, pp. 351--364, 2022.

\bibitem{chen2020devilSpeech2}
Y.~Chen, X.~Yuan, J.~Zhang, Y.~Zhao, S.~Zhang, K.~Chen, and X.~Wang, ``Devil’s whisper: A general approach for physical adversarial attacks against commercial black-box speech recognition devices,'' in \emph{29th USENIX Security Symposium (USENIX Security 20)}, 2020, pp. 2667--2684.

\bibitem{carlini2017towards}
N.~Carlini and D.~Wagner, ``Towards evaluating the robustness of neural networks,'' in \emph{2017 IEEE Symposium on Security and Privacy (SP)}.\hskip 1em plus 0.5em minus 0.4em\relax IEEE, 2017, pp. 39--57.

\bibitem{croce2021mind}
F.~Croce and M.~Hein, ``Mind the box: $ l\_1 $-apgd for sparse adversarial attacks on image classifiers,'' in \emph{International Conference on Machine Learning}.\hskip 1em plus 0.5em minus 0.4em\relax PMLR, 2021, pp. 2201--2211.

\bibitem{kereliuk2015deep}
C.~Kereliuk, B.~L. Sturm, and J.~Larsen, ``Deep learning and music adversaries,'' \emph{IEEE Transactions on Multimedia}, vol.~17, no.~11, pp. 2059--2071, 2015.

\bibitem{duan2022perception}
R.~Duan, Z.~Qu, S.~Zhao, L.~Ding, Y.~Liu, and Z.~Lu, ``Perception-aware attack: Creating adversarial music via reverse-engineering human perception,'' in \emph{Proceedings of the 2022 ACM SIGSAC conference on computer and communications security}, 2022, pp. 905--919.

\bibitem{rakamaric2014smack}
Z.~Rakamari{\'c} and M.~Emmi, ``{SMACK}: Decoupling source language details from verifier implementations,'' in \emph{Computer Aided Verification: 26th International Conference, CAV 2014, Held as Part of the Vienna Summer of Logic, VSL 2014, Vienna, Austria, July 18-22, 2014. Proceedings 26}.\hskip 1em plus 0.5em minus 0.4em\relax Springer, 2014, pp. 106--113.

\bibitem{luo2022frequency}
C.~Luo, Q.~Lin, W.~Xie, B.~Wu, J.~Xie, and L.~Shen, ``Frequency-driven imperceptible adversarial attack on semantic similarity,'' in \emph{2022 IEEE/CVF Conference on Computer Vision and Pattern Recognition (CVPR)}.\hskip 1em plus 0.5em minus 0.4em\relax IEEE Computer Society, 2022, pp. 15\,294--15\,303.

\bibitem{ali2023explainable}
S.~Ali, T.~Abuhmed, S.~El-Sappagh, K.~Muhammad, J.~M. Alonso-Moral, R.~Confalonieri, R.~Guidotti, J.~Del~Ser, N.~D{\'\i}az-Rodr{\'\i}guez, and F.~Herrera, ``Explainable artificial intelligence (xai): What we know and what is left to attain trustworthy artificial intelligence,'' \emph{Information Fusion}, vol.~99, p. 101805, 2023.

\bibitem{zhou2016learning}
B.~Zhou, A.~Khosla, A.~Lapedriza, A.~Oliva, and A.~Torralba, ``Learning deep features for discriminative localization,'' in \emph{Proceedings of the IEEE Conference on Computer Vision and Pattern Recognition}, 2016, pp. 2921--2929.

\bibitem{lin2013network}
M.~Lin, ``Network in network,'' \emph{arXiv preprint arXiv:1312.4400}, 2013.

\bibitem{selvaraju2017grad}
R.~R. Selvaraju, M.~Cogswell, A.~Das, R.~Vedantam, D.~Parikh, and D.~Batra, ``Grad-cam: Visual explanations from deep networks via gradient-based localization,'' in \emph{Proceedings of the IEEE International Conference on Computer Vision (ICCV)}, 2017, pp. 618--626.

\bibitem{oquab2015object}
M.~Oquab, L.~Bottou, I.~Laptev, and J.~Sivic, ``Is object localization for free?-weakly-supervised learning with convolutional neural networks,'' in \emph{Proceedings of the IEEE Conference on Computer Vision and Pattern Recognition (CVPR)}, 2015, pp. 685--694.

\bibitem{jacobgilpytorchcam}
J.~Gildenblat and contributors, ``Pytorch library for cam methods,'' \url{https://github.com/jacobgil/pytorch-grad-cam}, 2021.

\bibitem{marafioti2020gacela}
A.~Marafioti, P.~Majdak, N.~Holighaus, and N.~Perraudin, ``{GACELA}: A generative adversarial context encoder for long audio inpainting of music,'' \emph{IEEE Journal of Selected Topics in Signal Processing}, vol.~15, no.~1, pp. 120--131, 2020.

\bibitem{varelas2018comparative}
K.~Varelas, A.~Auger, D.~Brockhoff, N.~Hansen, O.~A. ElHara, Y.~Semet, R.~Kassab, and F.~Barbaresco, ``A comparative study of large-scale variants of cma-es,'' in \emph{Parallel Problem Solving from Nature--PPSN XV: 15th International Conference, Coimbra, Portugal, September 8--12, 2018, Proceedings, Part I 15}.\hskip 1em plus 0.5em minus 0.4em\relax Springer, 2018, pp. 3--15.

\bibitem{liu2023coverhunter}
F.~Liu, D.~Tuo, Y.~Xu, and X.~Han, ``Coverhunter: Cover song identification with refined attention and alignments,'' in \emph{2023 IEEE International Conference on Multimedia and Expo (ICME)}.\hskip 1em plus 0.5em minus 0.4em\relax IEEE, 2023, pp. 1080--1085.

\bibitem{xu2018key}
X.~Xu, X.~Chen, and D.~Yang, ``Key-invariant convolutional neural network toward efficient cover song identification,'' in \emph{2018 IEEE International Conference on Multimedia and Expo (ICME)}.\hskip 1em plus 0.5em minus 0.4em\relax IEEE, 2018, pp. 1--6.

\bibitem{sturm2013gtzan}
B.~L. Sturm, ``The {GTZAN} {D}ataset: Its contents, its faults, their effects on evaluation, and its future use,'' \emph{arXiv preprint arXiv:1306.1461}, 2013.

\bibitem{li2024mert}
Y.~Li, R.~Yuan, G.~Zhang, Y.~Ma, X.~Chen, H.~Yin, C.~Xiao, C.~Lin, A.~Ragni, E.~Benetos, N.~Gyenge, R.~Dannenberg, R.~Liu, W.~Chen, G.~Xia, Y.~Shi, W.~Huang, Z.~Wang, Y.~Guo, and J.~Fu, ``{MERT: Acoustic music understanding model with large-scale self-supervised training},'' in \emph{International Conference on Learning Representations (ICLR)}, 2024.

\bibitem{kilgour2019frechet}
K.~Kilgour, M.~Zuluaga, D.~Roblek, and M.~Sharifi, ``Fréchet {A}udio {D}istance: A reference-free metric for evaluating music enhancement algorithms,'' in \emph{Proceedings of Interspeech}, 2019.

\bibitem{gray1976distance}
A.~Gray and J.~Markel, ``Distance measures for speech processing,'' \emph{IEEE Transactions on Acoustics, Speech, and Signal Processing}, vol.~24, no.~5, pp. 380--391, 1976.

\bibitem{deng2020universal}
Y.~Deng and L.~J. Karam, ``Universal adversarial attack via enhanced projected gradient descent,'' in \emph{2020 IEEE International Conference on Image Processing (ICIP)}.\hskip 1em plus 0.5em minus 0.4em\relax IEEE, 2020, pp. 1241--1245.

\bibitem{wierstra2014natural}
D.~Wierstra, T.~Schaul, T.~Glasmachers, Y.~Sun, J.~Peters, and J.~Schmidhuber, ``Natural evolution strategies,'' \emph{The Journal of Machine Learning Research}, vol.~15, no.~1, pp. 949--980, 2014.

\bibitem{chen2017zoo}
P.-Y. Chen, H.~Zhang, Y.~Sharma, J.~Yi, and C.-J. Hsieh, ``Zoo: Zeroth order optimization based black-box attacks to deep neural networks without training substitute models,'' in \emph{Proceedings of the 10th ACM workshop on artificial intelligence and security}, 2017, pp. 15--26.

\end{thebibliography}

%
%
%
%

\end{document}